\documentclass[twocolumn]{svjour3}          
\smartqed  
\usepackage{xcolor}
\usepackage[switch]{lineno}

\usepackage{graphicx}
\usepackage{hyperref}

\journalname{Virtual Reality}
\begin{document}

\title{Trick the Body Trick the Mind: Avatar representation affects the perception of available action possibilities in Virtual Reality
}

\titlerunning{Avatar representation affects the perception of available action possibilities in Virtual Reality}

\author{Tugce Akkoc \and Emre Ugur \and Inci Ayhan }


\institute{T. Akkoc \at
              Bogazici University, Cognitive Science MA Program \\
              \email{akkoctugce@gmail.com}           
           \and
            E. Ugur \at Bogazici University, Computer Engineering Department
           \and 
           I. Ayhan \at Bogazici University, Psychology Department
}

\date{}

\maketitle

\begin{abstract}
In immersive Virtual Reality (VR), your brain can trick you into believing that your virtual hands are your real hands. Manipulating the representation of the body, namely the avatar, is a potentially powerful tool for the design of innovative interactive systems in VR. In this study, we investigated interactive behavior in VR by using the methods of experimental psychology. Objects with handles are known to potentiate the afforded action. Participants tend to respond faster when the handle is on the same side as the responding hand in  bi-manual speed response tasks. In the first experiment, we successfully replicated this affordance effect in a Virtual Reality (VR) setting. In the second experiment, we showed that the affordance effect was influenced by the avatar, which was manipulated by two different hand types: 1) hand models with full finger tracking that are able to grasp objects, and 2) capsule-shaped - fingerless- hand models that are not able to grasp objects. We found that less than 5 minutes of adaptation to an avatar, significantly altered the affordance perception. Counter intuitively, action planning was significantly shorter with the hand model that is not able to grasp. Possibly, fewer action possibilities provided an advantage in processing time. The presence of a handle speeded up the initiation of the hand movement but slowed down the action completion because of ongoing action planning. The results were examined from a multidisciplinary perspective and the design implications for VR applications were discussed.
\keywords{Virtual Reality \and Avatars \and Affordance \and Virtual Hands}
\end{abstract}

\section{Introduction}
\label{intro}

With recent breakthroughs in Virtual Reality technology, VR has the power to redefine the way we interact with digital media. Today, VR devices are low-cost and accessible. There is a growing number of researchers, developers, VR enthusiasts, and tech companies exploring and enhancing the user experience and interactivity in immersive virtual environments. Although much effort has been put on the exploration of the perception of virtual bodies and virtual environments in VR separately, little attention has been given to their interaction. 

Gibson (1966) coined the term affordance, which became a commonly used concept in explaining the interaction between an agent and its environment. The same object can afford different actions for different agents. A bottle, for example, is `grasp-able' for a person, rather `climb-able' for an ant. Thus, it is the relationship between the agent and the environment that determines the possible ways of interaction. Without technologies like VR, on the other hand, it is not easy to manipulate human body to observe the impact of morphological changes on perception.  In fact, the literature on affordance is heavily based on behavioral studies where bodily morphology remains unaltered, whereas the properties of the environment and/or objects are changed in various ways \cite{ellis2000micro,handy2003graspable,tipper2006vision,ren2013human,yamani2016object}.

Warren conducted the first empirical study to show that affordance perception is based on bodily perception by studying human stair climbing with short and tall people \cite{warren1984perceiving}. Warren and Whang provided further evidence that affordance perception is body-scaled by studying the passability of an aperture with large and small subjects \cite{warren1987visual}. However, body perception is not limited to innate differences in body morphology between individuals. Tool-use is known as a way of manipulating body perception. People identify tools as extensions of their own bodies \cite{serino2007extended,sposito2012extension}, which in turn changes the way they perceive their environment \cite{berti2000far,farne2005shaping,witt2005tool,canzoneri2013tool}. Just holding a stick, for example, can create a dramatic effect on our body perception such that it remaps our body form and the environment around us and the interactivity in between. It has been shown that tool-use as brief as 15 minutes is enough for people to overestimate their arm length \cite{sposito2012extension}. Through the interactions with the environment, the brain constantly updates the neural representation of the body parts and their positions \cite{sirigu1991multiple}. Therefore, humans do not perceive their bodies in stable metrics. Instead, our body perception is dynamic and changes through action. 

Both VR and tool-use studies provide evidence for the flexibility of body perception. It is possible to control altered bodies in VR through the manipulation of sensory feedback. As sensory feedback changes, our perception of our bodies and the world around us change, too. The perception of body and environment are interdependent (For a review, see \cite{harris2015our}). Yet measuring this interaction between the body and affordance perception requires controlled manipulation of the body form and its capabilities, which is a highly challenging task in experimental design. VR technology allows us to manipulate human body through the embodiment of avatars, and measure affordance perception while in action. 

Here, by using different avatar hands in a VR environment, we demonstrate – for the first time in literature - that reaction time to act upon an object is modulated differentially for restricted versus capable avatars and that this affordance effect depends on the phase of action, whether it is during action preparation or action execution. Affordance effect is measured by the change in response time according to the orientation of the graspable part of an object with respect to the response hand. Contrary to our expectation, when we removed the avatar's capability of grasping in the presence of a graspable object, we saw a reduction in the reaction time that indicates faster action planning. These results are in line with the affordance competition hypothesis \cite{cisek2007cortical} and will be further discussed in detail in the discussion section. Partitioning the response time measurement into different phases of action was a critical decision for our study. Strikingly, we found that the affordance effect shows a different pattern depending on the phase of the action. Furthermore, having fingers or not only matters during the planning time of a grasping action and creates no difference for the overall time to complete the action. Rosenbaum showed that planning occurs before the initiation  time \cite{rosenbaum1980human}.  Therefore, only in the planning time, we saw the effect of having a hand that can grasp while acting on a graspable object. Fleming et al. suggested that planning and preparation do not completely stop and continue during and even after action execution \cite{fleming2002time}. People adjust their hands and finger positions according to the object properties. In our study, we found that the existence of a handle accelerated the initiation of the hand movement but slowed down the completion of the movement because of ongoing action planning. Last but not least of all, we showed that less than 5 minutes of adaptation time in VR is enough to alter affordance perception. 

\subsection{Related Work}
\subsubsection{Redesigning the human body}
Avatars are 3D representations of the body and its movements in Virtual Reality (VR). They create the feeling that a virtual body is one's own and one has control over the movements of that body \cite{debarba2018self}. In immersive virtual environments, avatars provide a tool for manipulating the body and its capabilities. In a series of informal studies, Lanier et al. explored the limits of our ability to adapt to novel bodies in Virtual Reality (VR) by testing awkward avatars \cite{lanier2006homuncular}. For more than 10 years, they prototyped and tested avatars that are radically different from the standard human body form but still controllable. These studies have shown that people are good at adapting to avatars with different body forms to such an extent that they can even control a lobster body with eight arms. 

Mastering the use of a novel body and being able to control it skillfully in VR requires a certain level of embodiment. In this context, embodiment refers to the sense of ownership and agency created by the avatar. In other words, the avatar becomes ``the new source of sensations'' as if the actual body of the user is substituted by a virtual one \cite{gonzalez2018avatar}.

Avatars in immersive virtual environments are embodied through a first-person perspective, synchronous visuotactile, and/or sensory-motor feedback. Embodiment can occur with avatars of different ages \cite{banakou2013illusory}, gender \cite{fizek20114}, and race \cite{salmanowitz2018impact}. In fact, people can learn to control avatars with radically different characteristics than the standard human body form such as avatars with disproportionate body parts \cite{kilteni2012sense}, in the form of different species \cite{ahn2016experiencing}, or even abstract representations \cite{roth2016avatar}.

Seeing virtual body parts in isolation is enough to create a sense of ownership in the absence of full-body tracking. The idea of perceptually replacing body parts with artificial objects dates back to the well-known rubber hand experiment. The Rubber Hand Illusion \cite{botvinick1998rubber} is a phenomenon where people feel like a rubber hand is their own when they observe it is stroked in the same way as their hidden hand. Recently, Aldhous et al. replicated the rubber hand illusion in a VR setting \cite{aldhous2017digital}. In a setup where there has been a mismatch between the positions of the hidden real and visible virtual hands, participants were asked to close their eyes and nail the spot where they thought their hand was located. Bias in the responses was indicated by the proprioceptive drift, which is the deviation of the reported position from the original spot. Results demonstrated that the proprioceptive drift was towards the virtual hand, suggesting that participants rely more on the visual cues than on the proprioceptive sensation, although the latter clearly signals where their actual hand actually is. Therefore, the visual representation of the hand that the user relies on can potentially mislead the user about the position, form, or abilities of the hand. Passive observation of the changes in the body representation can create perceptual alterations on the position and form of the body parts. Understanding these changes in perceived abilities, however, requires experimental designs where a participant actively interacts with the virtual objects. 

Won et al. emphasized the difference between ``being versus doing in a novel body'' \cite{won2015homuncular}. In an experiment, they demonstrated that participants learned to control an avatar with three arms, with an extra arm coming out of their chest. Participants with a third arm performed better than those in the control group who had avatars with two arms in a virtual reach and touch task, where having a long third arm is advantageous. These results suggest that task requirements are crucial in adjusting to a novel avatar body. Thus, here, in order to ensure the embodiment of novel avatar hands, we implemented an adaptation procedure, where participants were asked to carry out simple tasks using a virtual object such as reaching, pushing, and picking before the experimental trials. 

Redesigning the human body means redesigning the capabilities of that specific body form in VR. For example, users can reach distant objects and grasp them with a ``magical'' hand, so that the distant objects can suddenly become graspable and within reach. Or ``ghost hands'' can pass through virtual objects so the objects that look and feel solid are no longer graspable. Affordances are thus highly flexible in VR. This idea brought up the main question behind our study: Can altered avatar abilities affect a VR user's perception of action possibilities, known as affordances, in VR? Simply, can putting a person in a new body change their belief in what they can or cannot do?

\subsubsection{Altered abilities altered perception}
Our actual or even perceived abilities with respect to the actions available to us in a given context play an important role in our perception of the environment. Older, heavier or shorter people, and women, for example, overestimate the steepness of a staircase \cite{eves2014there}. Fatigue or being low on physical fitness \cite{eves2014there} or wearing a heavy backpack \cite{bhalla1999visual} also alter perception - hills appear to be steeper to the participants. Similarly, with ankle weights, participants perceive gaps as wider than they physically are \cite{lessard2009look}. Therefore, changing the body form can change the perceived action possibilities in a given interactive virtual environment.

Within the context of affordance, ``the way living beings perceive the world is deeply influenced by the actions they are able to perform'' \cite{jamone2016affordances}. The available actions we perceive when looking at an object are dependent on the object features. The percept of these features, on the other hand, are not invariant as Turvey suggested \cite{turvey1992affordances} but rather relative to the size of our body (or body parts) with reference to the perceived eye-height (as an index of scale). Thus, there is no internal measurement of absolute metric values based on object geometry. As we interact with the objects in daily life, for example, we intrinsically decide if a chair is sittable \cite{mark1987biodynamic} or a stair is climbable \cite{warren1984perceiving} according to our leg length. Similarly, as the users interact with the environment in VR, their perception of object features such as distance, scale or orientation may change since they recalibrate their movements according to the relationship between the virtual objects and their virtual body-avatar. In fact, Banakou et al. provided evidence that adults inhabiting 4-year-old avatar bodies in VR feel strong body ownership and overestimate the size of virtual objects \cite{banakou2013illusory}. 

As the body form changes, the perception of the relationship between the body and an object like distance may change, say if a person has longer arms. In a study, where they tested the effect of extended arms on one's perceived action capabilities, Day et al. found that calibration to an avatar with long arms is possible as long as participants receive feedback on their actions \cite{day2019examining}. Day et al. also emphasized the difference between adaptation and calibration. Whereas the former occurs in a longer period of time (e.g. adapting to the use of a prosthetic arm), the latter occurs rather quickly (e.g. recalibrating responses according to the form and capabilities of an avatar body). Not only visual feedback changes one's perceived action capabilities, but perceived action capabilities, and thus, the embodiment of avatars can also affect the way we see and react to the world around us. Testing this hypothesis, on the other hand, requires a reliable measurement of affordance perception. In the next section, we provide a background for the selected methodology.    

\subsubsection{How to measure affordance perception in Virtual Reality}
Most studies examining avatar embodiment or interactivity in VR are in the form of formal/informal user studies based on user feedback, questionnaires, semi-structured interviews or observation \cite{lee2015transection,le2016giant,miura2016natural,ahn2016experiencing,pfeuffer2017gaze}. On the other hand, some embodiment studies manipulating the body morphology in VR are limited to quantitative changes like size \cite{slater2008towards} or the number of the limbs \cite{won2015homuncular}, the scale of the body \cite{banakou2013illusory}. In our study, we focused on changing the avatar functionally by giving participants normal (able to grasp) or fingerless hands (not able to grasp) while they are presented with a stimulus that has a graspable handle and we measured affordance perception with a robust methodology. 

Affordance perception is known to be formed in a short while, exerting its effects during action planning, even before an observer recognizes an object and makes a conscious decision on what to do with it. In a review of the affordance literature,  Jamone et al. provided three conclusive insights from a multidisciplinary perspective: ``(A) perception of action-related object properties is fast; (B) perception and action are tightly linked and share common representations; (C) object recognition and semantic reasoning are not required for affordance perception.'' \cite{jamone2016affordances}. If the process of action-related perception is fast, then the investigation should also be made by fast-paced measurement techniques. Within the context of perceptual-motor behavior, the fastest responses are also thought to be the most accurate \cite{fitts1953sr}).

Affordance perception can be measured with a fast-paced reaction time task by using the Stimulus-Response Compatibility (SRC) paradigm, where the compatibility between a stimulus feature and a given response in a task creates a measurable effect on the speed and/or accuracy of the response \cite{fitts1953sr}. For example, a circle can be presented on the right or the left side of the screen while a participant is responding according to the color of the stimulus. A participant may be responding with the right hand if the stimulus is red and responding with the left hand if the stimulus is blue regardless of its location.  This is a classic example of the Simon Effect \cite{simon1969reactions} that is based on the spatial relationship between the location of the stimulus and the response. Note that what is measured here (the compatibility between the stimulus position and the response) is independent of the task (left or right press contingent upon the color) which allows researchers to study the potentially unconscious effects of the stimulus on response time, which is also known as the SRC paradigm. This relationship between the stimulus and the response is called the compatibility effect. The same methodology can be used by changing the location of a part of the object (e.g. the handle), rather than the location of the whole object. For instance, the handle of a mug can be presented either on the right or left-hand side. When a stimulus is a signifier for action, then the compatibility effect is called the affordance effect \cite{ambrosecchia2015spatial}.

Instead of the color or shape of an object, Tucker and Ellis used object inversion as a criterion for the responding hand selection \cite{tucker1998relations}. In this task, participants responded with the right hand if the object was upright and with the left hand if it was upside down. In contrast with a color-based task, where the response depends on the color of the object, making a decision on the object orientation requires mental processing of the form, which carries information about the potential actions. Handled objects like a mug or a frying pan, for example, automatically potentiate a reach-and-grasp response towards their handles \cite{yamani2016object}. In a bimanual response task, left-hand responses are faster when the handle is on the left side, whereas right-hand responses are faster when the handle is on the right side \cite{cho2010object}.

Overall, this methodology allows us to measure the affordance effect while participants are engaged in an independent task and unaware of the main purpose of the experiment. There are also other advantages of this methodology in a VR setting. Firstly, the participant is isolated from the outside distractors and unable to see anything other than the experimental setup that is designed by the experimenter. This provides the experimental conditions to be precisely controlled. Secondly, the stimuli (3D models of the objects with a handle) is more realistic than the 2D pictures used in prior studies. Thirdly, the responses like grasping and reaching can be measured during both action preparation and action execution by dividing the measurement into subcategories of action. Thus, here, using a stimulus-response compatibility paradigm in a VR setting, we investigated the effect of the avatar representation on affordance perception. 

\subsection{Significance of the current study}
In this study, we present a robust methodology to measure interactivity in VR by quantifying the perceived action possibilities. In the first experiment, in a virtual room environment, participants were presented with a mug on a table that appeared either in an upright or upside-down orientation across different trials. The task of the participants was to respond as quickly as possible by rotating their wrist and making a grasp action using their left (i.e., for upside down decisions) or right (i.e., for upright decisions) hands. This compatibility between the reacted hand and the handle orientation was the first independent variable. In blocked trials, the distance between the participant and the mug was also manipulated - as a second independent variable - such that the mug could appear in 3 different locations - at a near, middle or far distance from the participant. The response times of the participants were recorded from the moment when the stimulus had first appeared until the grasping response was finalized. The main purpose of the first experiment was to replicate the affordance effect in a VR setting and the methodology was found to be effective. In the second experiment, the mug stimulus was replaced by a frying pan, the position of which was fixed at a single coordinate (the middle position used in the first experiment). In two different conditions of the independent variable, the avatar hand was either a realistic hand with both grasping and pushing abilities or a capsule-like restricted hand, which lacked any grasping abilities. Response times were recorded using two indices, namely the lift-off and movement times. Lift-off time was the time it took for participants to lift their hands off a rest-state-key on a keyboard following the presentation of the stimulus on the screen. Movement time, on the other hand, was recorded from the lift-off time until the hand model contacted the invisible, virtual detection box on the target location. Recording the lift-off time, as well as the movement time increased the reliability of the reaction time measurement. The results revealed different patterns for different phases of action. 

\section{Methods and Results}
\subsection{Experiment 1: Using Stimulus-Response Compatibility (SRC) paradigm to measure the affordance effect in VR}

The main purpose of this experiment was to validate our experimental setup and methodology. We aimed to test if we can create an affordance effect with a virtual object in a virtual environment when participants only see virtual hands mimicking the movements of their actual hands, which are out of sight. Objects with handles are known to potentiate afforded action. Participants tend to respond faster when the handle is on the same side as the responding hand in a bimanual speed response task \cite{tucker1998relations}. Additionally, this effect, also known as the affordance effect, is found to be affected by the reachability of the object \cite{costantini2010does}.

In the literature, distance is usually divided into two categories: peripersonal -within reach- and extrapersonal space -out of reach- \cite{costantini2011objects,costantini2010does,ferri2011objects}. In this experiment, we also introduced a mid-range position between the two in order to see whether the affordance effect is categorical (i.e. reaction times clustering around the two categories — the effect is present or absent) or gradient (the strength of the effect varies gradually).

{\bf H1:} {\em On a SRC paradigm measuring the affordance effect in VR, participants will respond faster when the object handle is on the same side as the responding hand than when the object handle is on the opposite side.}

{\bf H2:} {\em The SRC effect in the virtual environment will only be found when the object is at a reachable distance (in the peripersonal space) than when it is at an out-of-reach distance (in the extrapersonal space) leading to a categorical rather than a gradual change.}

\begin{figure*}
\centering
\includegraphics[width=0.75\textwidth]{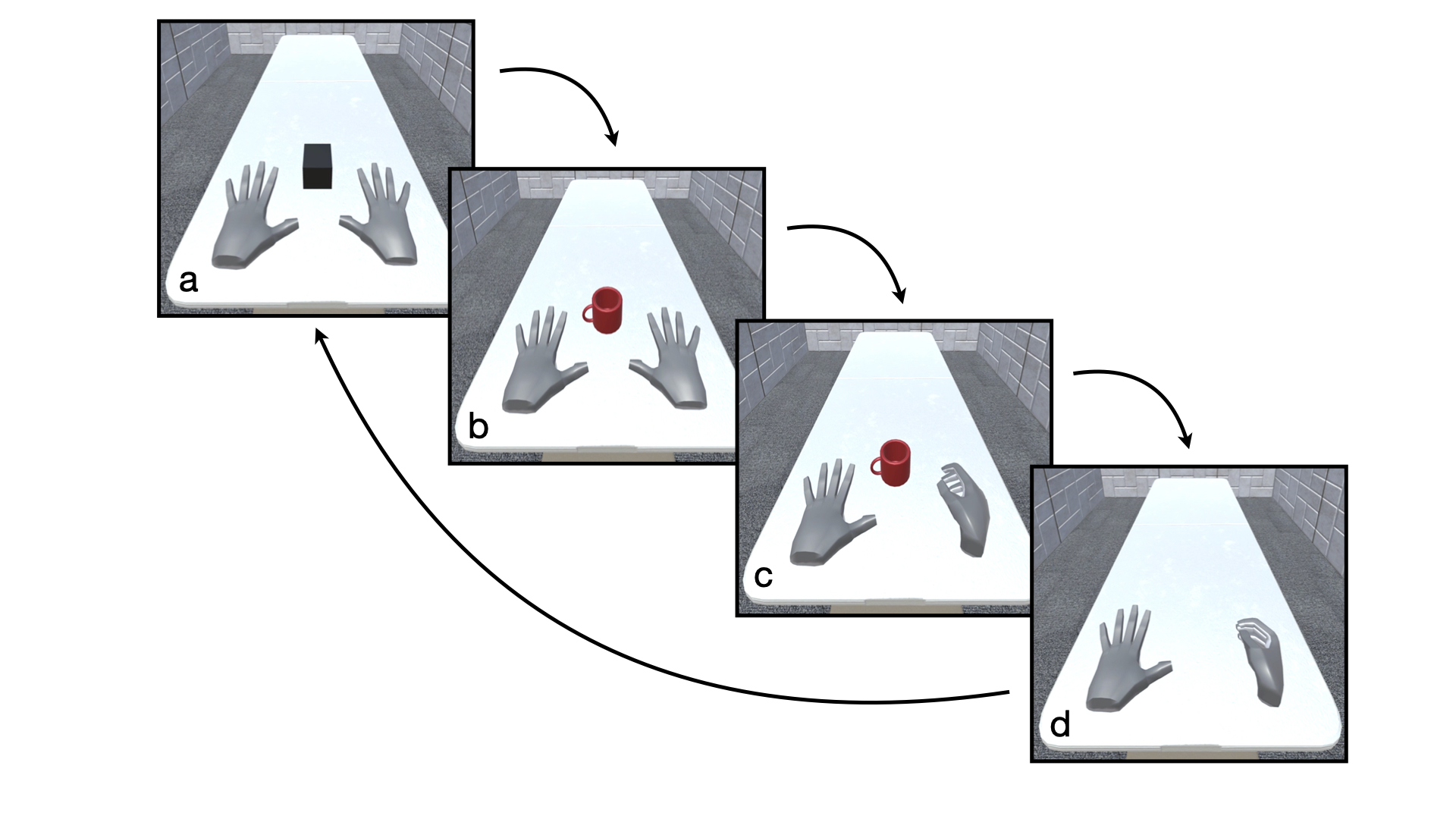}
\caption{Outline of a trial in Experiment 1.a) At the beginning of each trial, a black cube appeared in one of the three locations on the long table (Near, Middle, Far). b) After 500 milliseconds, the cube disappeared and the stimulus randomly appeared at the same location in one of four different orientations (Right-Up, Right-Down, Left-Up, or Left-Down). Up and Down represent the vertical orientation of the stimulus, while Right and Left represent the handle orientation. The handles were rotated 15$^{\circ}$ towards the participant. c) Participants responded according to the vertical orientation of the stimulus by making a grasping gesture either with their right or left hand.  d) At the end of the grasping gesture, the target disappeared and the trial ended. See the video of the experimental procedure: \url{https://youtu.be/7vc_zBAQxMM}} 
\label{fig:outline1}       
\end{figure*}

\subsubsection{Methods and Materials}
\paragraph{Participants:} Sixteen graduate and undergraduate students (13 male and 3 female) from Istanbul Technical University participated in Experiment 1. The mean age of the participants was 25 ranging from 22 to 33 years old. Fifty-six percent of the subjects had previously tried VR at least once. Seventy-five percent of the participants were right-handed and the rest were left-handed. The study was compliant with the Bogazici University research ethics requirements, as well as the Declaration of Helsinki.

\paragraph{Apparatus:} Virtual hands are usually tracked by VR controllers. The position and rotation of the controllers are mapped onto the virtual representation of the hands. Here, we used the Leap Motion hand tracking device\footnote{\url{https://www.ultraleap.com/}} so that the movement of the participant's bare hands and fingers could be tracked and modeled in VR. The hand tracking device was connected to a laptop (HP OMEN) and physically attached to the front of a head-mounted display (Acer Windows Mixed Reality Headset\footnote{\url{https://www.acer.com/ac/en/US/content/series/wmr
}}). Participants saw the stimulus via a head-mounted display. Since the VR goggles physically covered their eyes, they did not see the real experimental room but rather the virtual room from a first-person perspective. In both the real and virtual experimental rooms, participants were sitting on a chair in front of a table. A Sony 310AP Wired Headphones was connected to the VR headset for auditory feedback. A red mug with a handle on the right-hand or left-hand side was used as a stimulus that randomly appeared in different orientations (upright or inverted). In blocked trials, the mug was presented either at a Near (30 cm), Middle (60 cm), or Far (150 cm) distance from the participant.  

\paragraph{Procedure:} First, the experimenter demonstrated the resting position of the hands. In the default resting position, participants held both of their hands open (in front of the tracking device) with their palms facing away from them. They were then shown the required hand movement to respond to the stimulus: rotating the wrist inward and grasping. The experimenter then presented the upright and inverted orientations of the stimulus with a real physical red mug which was similar to the virtual mug (which acted as the stimulus) they were about to see in the virtual environment.

Participants initiated the experiment by pressing the space bar on a keyboard. They were instructed to fixate their eyes on a black cube (with the side lengths of 10 cm) that appeared in one of the three positions (at the Near, Middle or Far distance) for 500 milliseconds. Right after the cube disappeared, a red mug appeared at the same position as the black cube. The task of the participants was to make a grasping gesture using the correct hand according to the vertical orientation of the mug (whether upright or upside down), independent of the handle orientation (Fig.~\ref{fig:outline1}). Following each grasp response was faint auditory feedback to mark the end of an individual trial. This stereo sound always originated from the same location (left or right) of the responding hand. After every trial, participants returned their hands back to their default position.  Response times were automatically recorded during the experiment. For every individual trial, the response time counter started at the appearance of the object and ended when the thumb and index finger touched each other in order to complete the grasping gesture.

For the first part of the experiment, participants completed a practice session with 25 trials during which they were given auditory error feedback. Participants who made errors in more than 10\% of the trials repeated the practice session. At the beginning of the experimental session, participants were instructed to respond as quickly and accurately as possible. Each experimental block was composed of 60 test trials. Half of the participants responded with the right hand when the object is upright and with the left hand when the object is upside down in the first block, and responded with the left hand when the object is upright and with the right hand when it is upside down in the second block. The mapping of response hand to object inversion was changed and counterbalanced across participants. Before the second block, the participants completed another 25 practice trials with the new instruction. Afterwards, they completed  60 test trials. Participants thus completed 120 trials in two blocks. Each test trial took approximately two minutes. In each block, the mug appeared at every combination of position and orientation (right-up, right-down, left-up, left-down) for 5 times. The order of these states was randomized.

\subsubsection{Results and Discussion}
\begin{figure}
\includegraphics[width=0.5\textwidth]{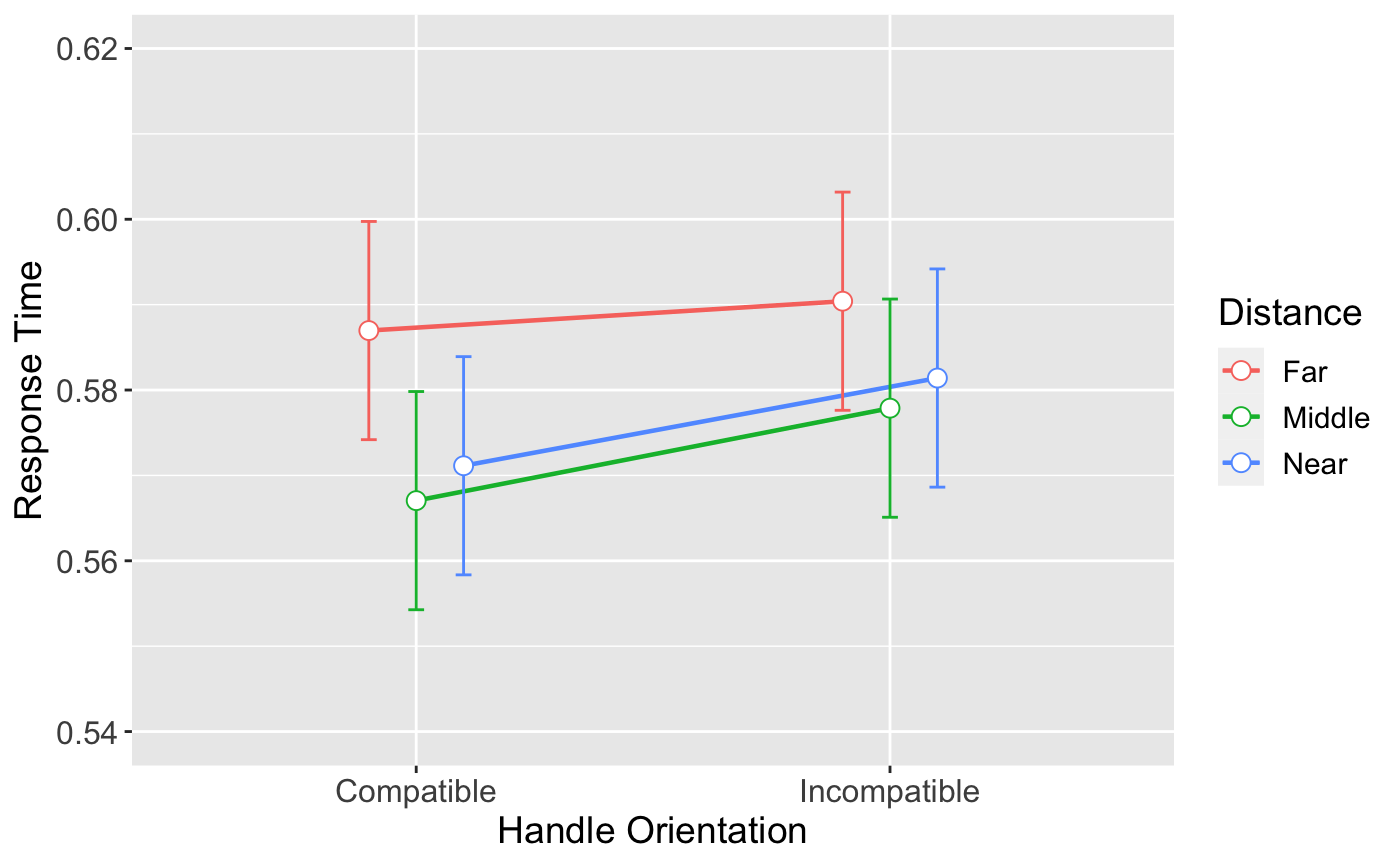}
\caption{Overall results of Experiment 1. Response time is represented on the y-axis in seconds. On the x-axis, two levels of the handle orientation effect (the compatible and incompatible conditions) are shown. The purple, orange, and green lines represent the conditions where the stimulus was shown at near, middle, and far distances respectively. Error bars indicate the standard errors of the mean (+/- 1 SEM). }
\label{fig:2}       
\end{figure}

Error trials and response times 2 standard deviation above or below from the condition means were excluded from the analysis. A 2 x 3 repeated measures analysis of variance (ANOVA) was conducted with the Handle Orientation (compatible and incompatible) and Distance (Near, Middle, and Far) as within-subject factors. The main effects for both Handle Orientation and Distance were significant, $F(1,15)=7.690, p=.014, \eta^2=.339$ and $F(2,15)=7.290,  p=.003, \eta^2=.327$, respectively. Participants reacted faster in the compatible conditions (Mean = .575 , SD  = .048) than in the incompatible conditions (Mean = .583, SD = .052).  A Tukey's HSD post hoc analysis on the main distance effect showed that participants responded significantly faster when the object is presented in the Near (Mean = .576, SD = .048) or Middle (Mean = .572, SD = .051) positions compared to the Far (Mean = .588, SD = .052) position. The difference between Near and Middle conditions were not significant. The interaction effect between Handle Orientation and Distance was found to be insignificant, $F(1,31)=0.435, p=.651, \eta^2=.028$ (Fig.~\ref{fig:2}).

These results provide further evidence for the findings of \cite{costantini2010does}, where they showed the main effects of handle orientation and distance of the object on reaction time in a 3D stereoscopic viewing setup combined with colored hand pictures. That we found a significant handle orientation effect in VR with hand tracking demonstrates that the stimulus-response compatibility paradigm can also be used to understand the interaction between an object and a virtual hand.

Although we could not find a significant interaction effect between handle orientation and distance, there has been a clear trend in the data. In the compatible conditions, participants reacted faster only when the object was in a reachable region category (Near and Middle) than when it was in the out-of-reach (Far) region (Fig.~\ref{fig:2}).  This is compatible with the Constantini et al. findings\cite{costantini2010does}, where object features were found to evoke actions only when presented in the peripersonal space. 

In summary, in Experiment 1, we showed that the affordance effect can be created in VR. In Experiment 2, we used this methodology to look at the effect of avatar representation on how people perceive the available action possibilities.

\subsection{Experiment 2: Manipulating the virtual hand representation}
The purpose of the second experiment was to see if the affordance effect is modulated by the body representation. Previous studies showed that avatar abilities can be changed in VR \cite{won2015homuncular,day2019examining}. We used two different hand types: one with fully animated fingers and one without fingers. The former is the able hand that it is able to grasp virtual objects, and the latter is the restricted hand that it is not able to grasp objects.  Using the same SRC paradigm that we tested in Experiment 1, here, in Experiment 2, we manipulated the avatar hand representation to measure the affordance effect. 

{\bf H3:} {\em On a SRC paradigm measuring the affordance effect in VR, participants will respond faster when the object handle is on the same side as the responding hand than when the object handle is on the opposite side.}
{\bf H4:} {\em The affordance effect in VR will be affected by the avatar hand type such that reaction times will differ for the restricted hand avatars than for the able hand avatars.}

\subsubsection{Methods and Materials}
\paragraph{Participants:} Thirty-two undergraduate students (14 female, 18 male) from Boğaziçi University participated in Experiment 2. The mean age of the participants was 23 ranging from 19 to 35 years old. Sixty-two percent of the subjects had previously tried VR at least once. Ninety-four percent of the participants were right-handed and the rest were left-handed. None of the participants had participated in Experiment 1. The study was compliant with the Bogazici University research ethics requirements, as well as the Declaration of Helsinki.

\begin{figure}
\includegraphics[width=0.45\textwidth]{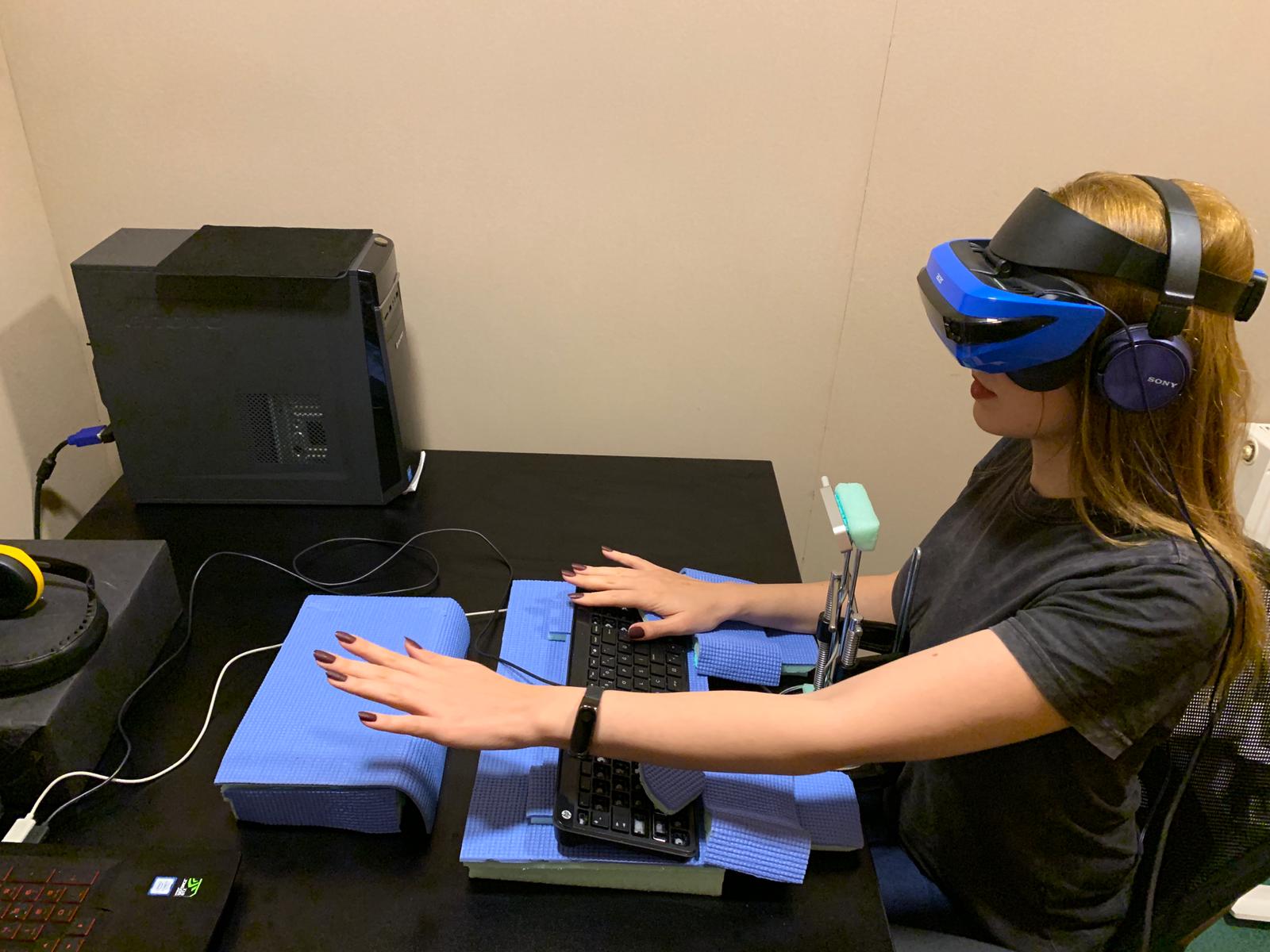}
\caption{Experimental setup for experiment 2. `X' and `2' on the numeric keypad were used as response keys. All the other keys surrounding them were pulled out of the keyboard to prevent possible interfering inputs during the experiment. The two response keys were physically enlarged. The keyboard was raised and aligned with the handrests. The Leap Motion hand tracking device was fixed at the edge of the table. The response was given by lifting the responding hand off the response key and reaching forward. 
}
\label{fig:3}       
\end{figure}

\paragraph{Apparatus:} The methodology was the same as the first experiment whereas the experimental setup and the measurement technique were changed  (Fig.~\ref{fig:3}). In this experiment, the hand tracking device was fixed at the edge of the table rather than the VR headset to prevent the errors that might be introduced by the reaching movements. An extra keyboard was connected to the same computer to divide response time measurement into two parts: lift-off time and movement time. Whereas the lift-off time indicated the interval between the appearance of the target stimulus and the release of the response key, movement time indicated the interval between the release of the response key and the touch over the target stimulus.

A frying pan, designed in Autodesk Maya 2018\footnote{https://www.autodesk.com/products/maya/}, was used as the stimulus object. The bottom texture of the pan was chosen so as to be different from that of the top part to make it easier for participants to discriminate the upright oriented pan from an inverted pan. Object positioning was done using the base-centered approach \cite{cho2010object} as cited in \cite{kostov2017handle} where the pivot of the virtual object was placed at the center of the body of the pan.
            
The distance of the pan from the participant was fixed at the `middle' distance used in Experiment 1. To the right and left sides of the pan were two invisible boxes to detect the contact of the hand models where the reaching response was completed. Two audio source objects were placed in the same locations for audio feedback to mark a successful reaching response.

\begin{figure}
\includegraphics[width=0.6\textwidth]{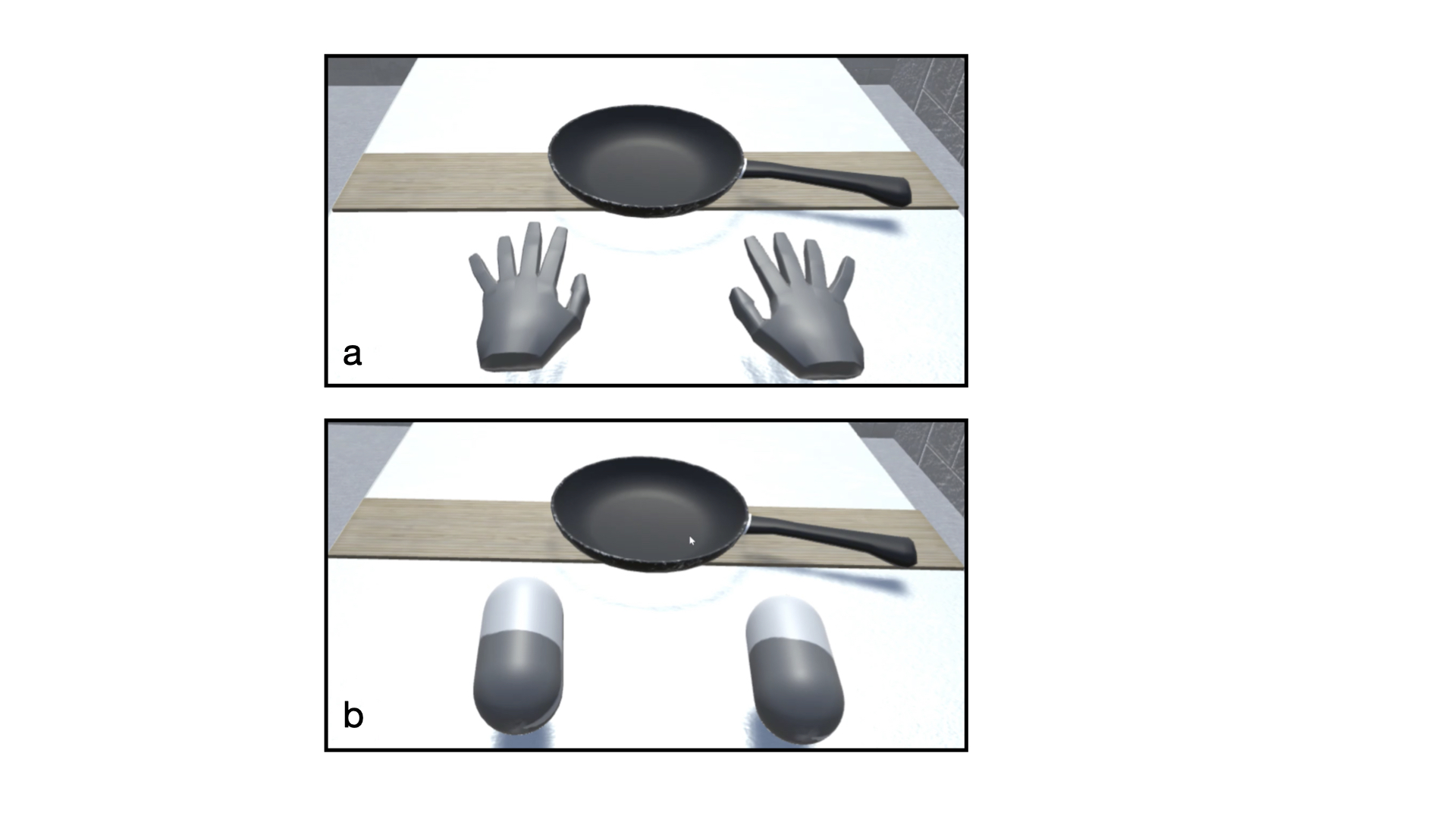}
\caption{Virtual hand representations. a) The able hand and b) the restricted hand.}
\label{fig:4}       
\end{figure}

The main difference between Experiment 1 and Experiment 2 was the second independent variable - avatar hand type. Two types of hands were used: Able and Restricted hand. The Able hand model was a full finger tracking hand model which was also used in Experiment 1 (Fig.~\ref{fig:4}a), while the Restricted model was a capsule or a pill-shaped 3D model (Fig.~\ref{fig:4}b). The restricted hand had the same size, texture, and color as the normal hand. The able hand was designed to allow grasping behavior as opposed to the restricted hand, which was not able to grasp. Thus, two avatar hands allowed different abilities to the participants. There were no physical restrictions on the real hand movements of the participants, but when they grasped with their real hands, only the full hand model provided visual feedback of the grasping action. Likewise, the Restricted hand lacked a grasping ability and simply collided with the virtual objects.

\begin{figure*}
\centering
\includegraphics[width=0.8\textwidth]{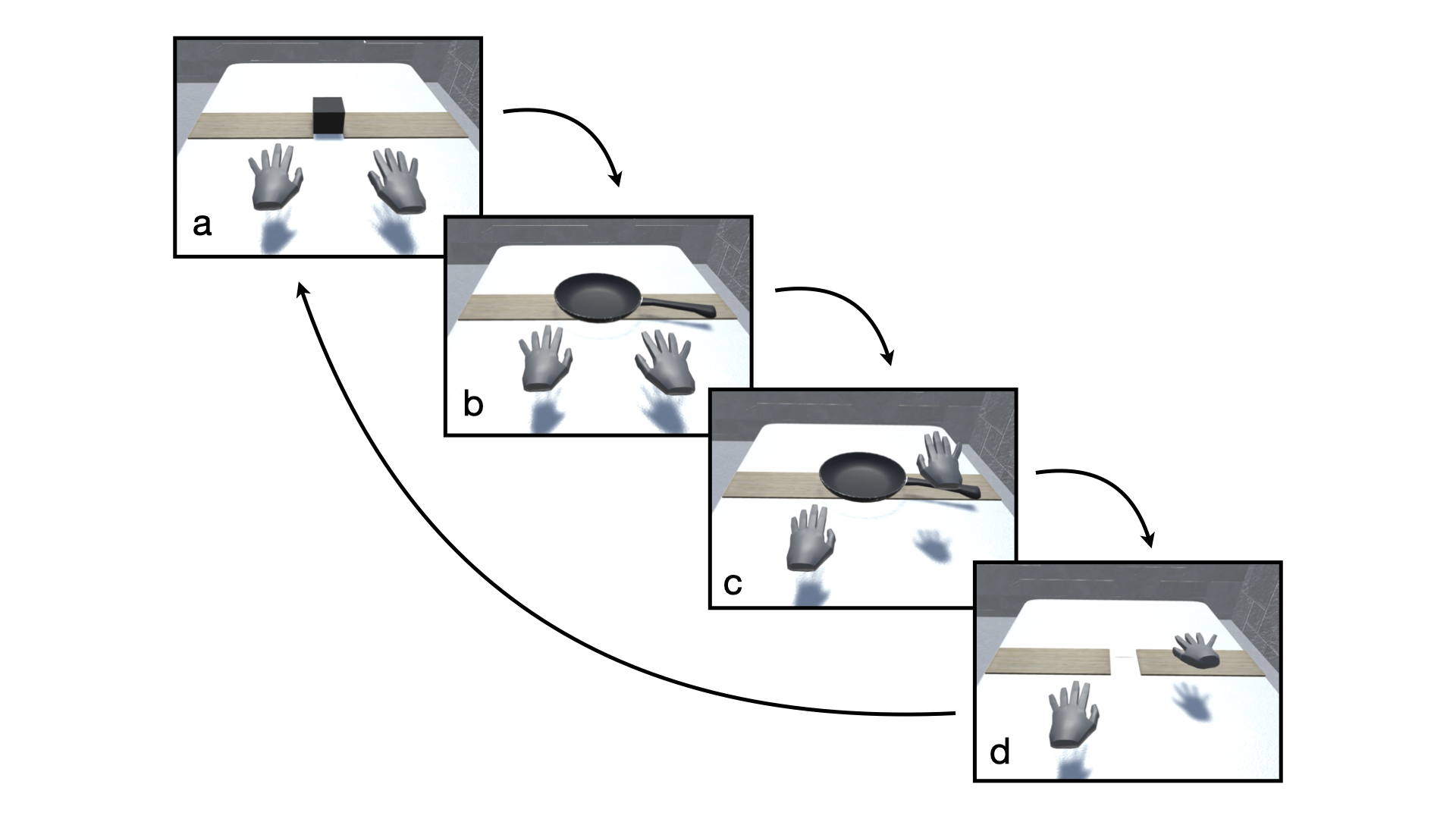}
\caption{Outline of a trial in Experiment 2. a) At the beginning of each trial, a black cube appeared at the same location on the table. b) After 500 milliseconds, the cube disappeared and the stimulus appeared at the same location in one of four different orientations randomly (Right-Up, Right- Down, Left-Up, or Left-Down). Up and Down represent the vertical orientation of the stimulus, while Right and Left refer to the handle orientation. The handles were rotated 15$^{\circ}$ towards the participant. c) Participants responded according to the vertical orientation of the stimulus by lifting their right or left hand off the response key and reaching forward until the target point, represented by the wooden areas on the table. d) As the virtual hand contacted the invisible detection box, the target disappeared and the trial ended. Please see the video of experimental procedure: \url{https://youtu.be/eyru1BVJShU}}
\label{fig:5}       
\end{figure*}
\subsubsection{Procedure}

Prior to each experimental session, participants spent time with the respective hand model for adaptation. In the adaptation phase, they performed 4 sessions composed of simple tasks as instructed by the experimenter 16 times each. Adaptation Task 1A was ``Open and Close Hands A''. With palms facing inwards participants opened and closed their hands while looking at their palms and counting out loud. Adaptation Task 1B was ``Open and Close Hands B''. With palms facing outwards participants opened and closed their hands 16 times while looking at the back of their hands and counting out loud. Adaptation Task 2 was ``Reach and Touch''.  Participants reached and touched red cubes that appeared either on the right-hand side or left-hand side with the corresponding hand and then brought their hand back to the default position. Adaptation Task 3 was ``Push''. Participants pushed the cubes forward with the back of their hands with a flick of the wrist. Adaptation Task 4 was ``Pick and Place''. Participants picked the red cubes up and placed them on black cubes.

In the experimental phase, participants initiated the experiment by pressing both response keys simultaneously. In a manner similar to Experiment 1, as both keys were pressed down, a black cube appeared for 500 ms. Participants were told to fixate their eyes on the position of the black cube. Right after the fixation object disappeared, the stimulus (the frying pan) appeared with the handle either on the right or the left-hand side. Participants' task was to respond according to the vertical orientation of the object (whether it was upright or upside down) (Fig.~\ref{fig:5}). Responses were given by lifting the responding hand off the response key and reaching forward (Fig.~\ref{fig:3}).

The instruction was counterbalanced between subjects. Half of the participants were given instruction A only, which was `Respond with your right hand if the pan is upright and respond with your left hand if the pan is upside down' and the other half of the participants received the instruction B only, which is `Respond with your right hand if the pan is upside down and respond with your left hand if the pan is upright'.

Participants first completed 24 practice trials with error feedback as in Experiment 1. Once the practice session was completed, they started the first test session, which was composed of 24 trials without error feedback. Participants completed 6 blocks of 24 trials. In each block, the pan appeared at every location, and in every orientation (right-up, right-down, left-up, left-down) twice. The order of these states was randomized. Overall the experiment had 7 sessions: one practice and six test sessions in two parts composed of three sessions for each hand type. Prior to each test session was an adaptation phase.  Depending on the pace of the participant, each adaptation phase took three to five minutes. Each test session took approximately one and a half minutes. Between the two parts of the experiment, participants were allowed to rest as long as they needed it. They were allowed to talk to the experimenter, and drink water but they were not permitted to have screen time or engage in cognitively or physically demanding tasks. The order of hand type was counterbalanced among the participants. After each session, data including lift-off time and movement-time for each trial was automatically saved. 

Immediately after the completion of the experiment, participants filled out two Body Ownership Questionnaires, one for Able hand (see Appendix A) and one for Restricted hand (see Appendix B). The questionnaire was prepared by modifying the questions that were previously used in three different VR studies \cite{argelaguet2016role,day2019examining,sikstrom2014role}. The questionnaires were given in the same order of the given hand type. Finally, participants were debriefed.

\begin{figure}[t]
\includegraphics[width=0.5\textwidth]{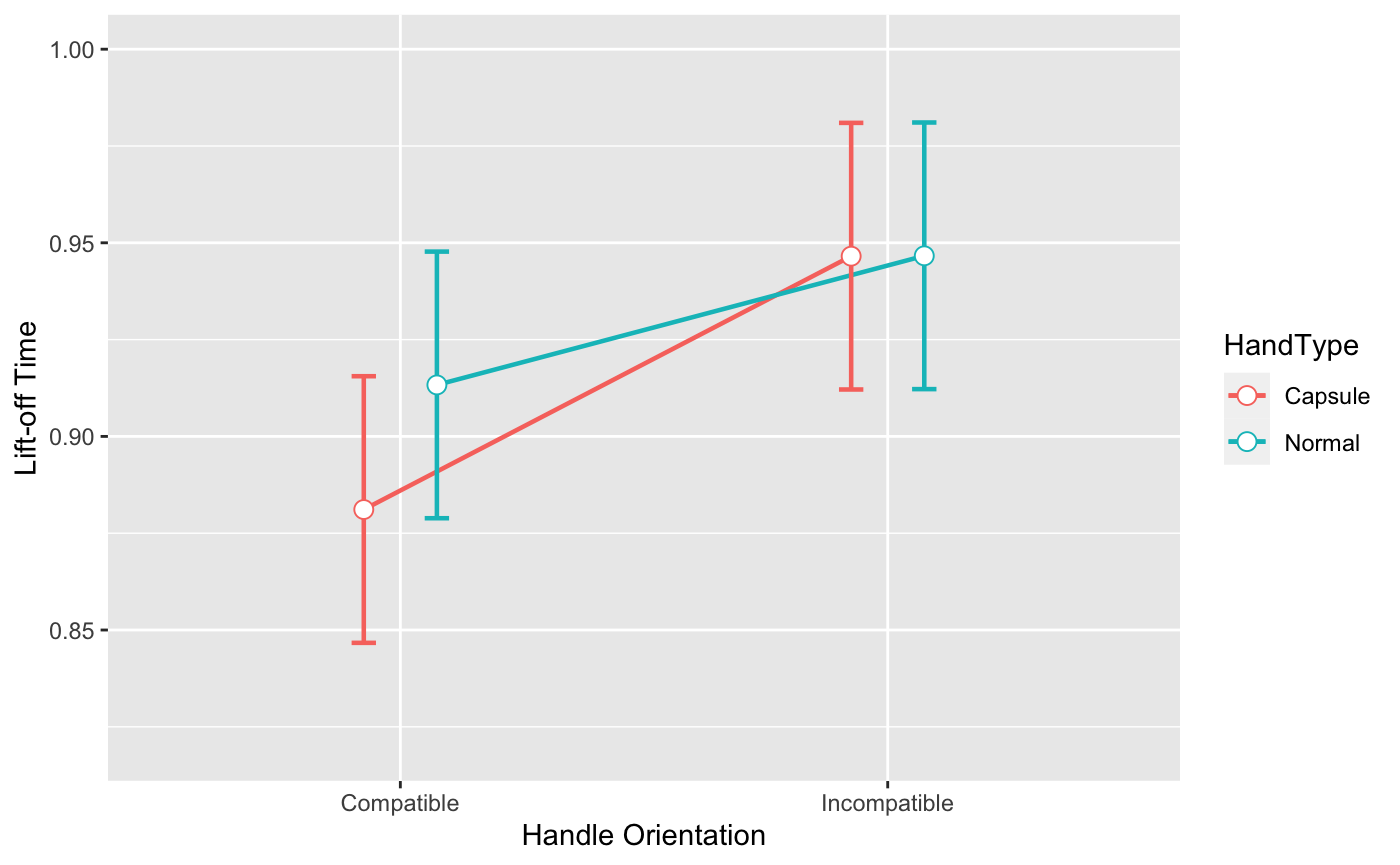}
\caption{Results of experiment two (lift-off time). Lift-off time is represented on the y-axis in seconds. On the x-axis, two levels of the handle orientation effect (the compatible and incompatible conditions) are shown. The green and orange lines represent the conditions where the participants have able or restricted hands respectively. Error bars indicate the standard errors of the mean (+/-1 SEM).}
\label{fig:6}       
\end{figure}

\subsubsection{Results}
For two different dependent variables (Lift-off Time and Movement Time), two Repeated measures analysis of variance (ANOVA) were conducted with the Hand Type (Able vs. Restricted) and Handle Orientation (compatible vs. incompatible) as within-subject factors.

{\bf Lift-off time} was the interval between the appearance of the target stimulus and the release of the response key. Error trials and response times 2 standard deviation above or below from the condition means were excluded from the analysis. A 2 x 2 within-subjects ANOVA indicated a significant handle orientation effect, $F(1,31)=34.456,  p=.001, \eta^2=.547$ such that participants lifted their hand off the response key faster in the compatible conditions (when the handle is on the same side as the responding hand) (Mean = .897, SD = .204) than in incompatible conditions (where the handle is on the opposite side of the responding hand) (Mean = .947 , SD = .183). There was no significant main effect of hand type, $F(1,31)=0.769, p=.387, \eta^2=.024$ which means that participants lifted their hands off the response keys at similar times with both able (Mean = .930, SD = .221) and restricted hand (Mean = .914, SD = .165).

More interestingly, there was a significant interaction effect between Handle Orientation and Hand Type, $F(1,31)=7.129, p=.012, \eta^2=.187$, demonstrating that the handle orientation effect was modulated by the hand type. As shown in Fig.~\ref{fig:6}, in the incompatible conditions, it took a similar amount of time for participants to respond using able (Mean = .947, SD = .203)  and restricted hand (Mean = .947, SD = .162 ). However, participants reacted significantly faster with the restricted hand (Mean = .881, SD = .164) than with the able hand (Mean = .913, SD = .239) in the compatible conditions.

\begin{figure}[t]
\includegraphics[width=0.5\textwidth]{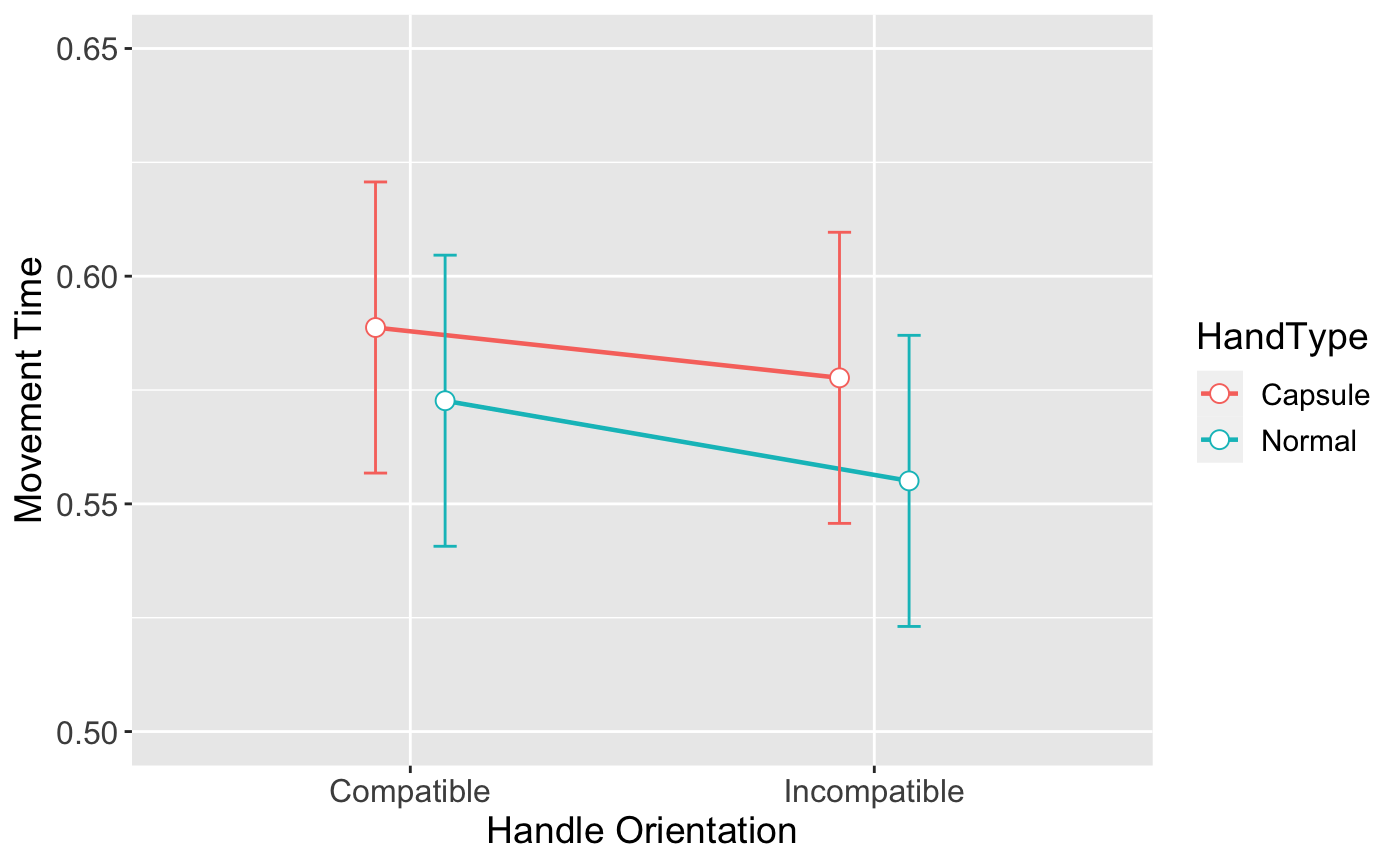}
\caption{Results of experiment two (movement time). Movement time is represented on the y-axis in seconds. On the x-axis, two levels of the handle orientation effect (the compatible and incompatible conditions) are shown. The green and orange lines represent the conditions where the participants have able or restricted hands respectively. Error bars indicate the standard errors of the mean (+/- 1 SEM).}
\label{fig:7}       
\end{figure}

{\bf Movement time} was the interval between the release of the response key and the touch over the target stimulus. Error trials and response times 2 standard deviation above or below from the condition means were excluded from the analysis. As well as in the lift-off time, handle orientation was also significant for movement time, $F(1,31)=9.414, p=.004, \eta^2=.233$, but with the reverse trend that it took longer to respond in the compatible trials (Mean = .581, SD = .178) than in the incompatible trials (Mean = .566, SD = .181) (Fig.~\ref{fig:7}). 

As in the lift-off time, the main effect of hand type was not statistically different for movement time, either, $F(1,31)=1.561, p=.221, \eta^2=.048$, thus, it took similar time to respond with the able (Mean = .564, SD = .171) and restricted hand (Mean = .583, SD = .188). Similarly, the interaction effect between the handle orientation and the hand type was also statistically insignificant, $F(1,31)=0.572, p=.455, \eta^2=.018$.

{\bf Body ownership} questionnaire included questions such as ``I felt as if the virtual hands that I saw during the experiment were part of my body'' to measure to what extent virtual body parts felt like one's own. Participants answered the same questions for both the able and restricted hand on a Likert scale from 0 to 10 corresponding to `completely disagree' and `completely agree'. The results demonstrated that both restricted and able hand provided a sense of ownership, although a paired-samples t-test, conducted to compare the sense of ownership created by different hand types, indicated that the able hand with full finger tracking created significantly more ownership (M = 7.99, SD = 1.14) than the restricted hand (M = 5.67,  SD = 1.74) condition, t(31) = 7.70,  p = .001,  d = 1.36. Since participants answered the questionnaires at the end of the experiment, however, they may have thought that they were supposed to compare the two hands and answer questions relative to one another. Thus, this difference between the hand types may not have appeared if it were a between-subject design, where participants would complete the task with only one hand type. 

Although participants reported that the able hand felt more like their own, however, the movement time results did not reveal a difference between the two hand types. Similarly, the lift-off time results did not reveal a difference between the able and restricted hands in the incompatible conditions, either. The only difference appeared in the compatible conditions, where the hand was approaching towards the handle of the object, implying that the effect of hand type is related to the relationship between the virtual hand's capability (able to grasp or not able to grasp) and object affordance (graspability) rather than the sense of ownership created by the hand type. In fact, in order to see whether the level of ownership has an effect on the significant interaction result in the lift-off time measures, we conducted a median-split analysis, where participants were divided into two groups (the ones who felt strong ownership and the ones who felt less strong ownership). An ANOVA analysis with the level of ownership given as a between-subject factor revealed that the effect of the level of ownership is insignificant, $F(1, 30)=3.03, p=.092, \eta^2=.092$, indicating that the ownership cannot explain the differences in the lift-off time.

\section{Discussion}

The findings of the experiments presented significant outcomes: 
\begin{enumerate}
\item Virtual objects can potentiate affordance effect in VR such that
		participants tend to respond faster when the handle of a target object is on
				the same side as the responding hand in a bimanual speed response task (Experiment 1 and 2). This affordance effect, on the other hand, depends on the phase of action: Whereas the trend is as described during the action planning, it is reversed during the action execution (Experiment 2).
\item The effect of avatar hand type on affordance perception also depends on the phase of action. Whereas the lift-off time is faster for the restricted hand avatar, the movement time is faster for the able hand avatar (Experiment 2).
\item The restricted avatar hand potentiates greater affordance effect than the able avatar hand (Experiment 2).
\item Adaptation to a new avatar hand is fairly fast in VR. Approximately a 5-minutes-adaptation which involves various interactions with objects is enough to embody a new body part.        
\item Reaction times in a bimanual speed response task are faster for virtually closer targets than for further targets (Experiment 1).
\item Having radically different hand designs in VR did not show any significant difference in the overall reaction time. Only the action planning time was different for different hand models (Experiment 2).
\end{enumerate}

Ecological approach to vision asserts that the environment is perceived in terms of behavioral affordances defined as what the environment offers to the agent. In this context, vision may be seen as an active exploration of external world mediated by the sensorimotor contingencies \cite{gibson1978ecological,noe2004action,o2001sensorimotor} as well as on-line guidance while executing actions. How these action possibilities inherent in the visual objects are represented in the brain, however, is a question yet to be answered. In a plausible account, the percept of an object automatically triggers the activation of motor representations of a list of potential competing actions (see \cite{cisek2007cortical}) even when the observer has no explicit intention to act \cite{tucker1998relations}.  These potentiated actions, on the other hand, depends on the particular state of a perceiver with respect to the object: A reach-and-grasp act, for example, cannot be executed if the object is beyond the reaching distance \cite{costantini2011objects}. Consistently with the previous findings, we hereby confirmed the stimulus-response compatibility effect – this time - in a virtual reality environment, where participants interacted with the object using a virtual hand, together with further evidence that the effect size is not gradual across the peripersonal towards the extrapersonal space but rather categorical. Finding a significant handle orientation effect in VR with hand tracking supported the idea that the stimulus-response compatibility paradigm can be used to understand the interaction between the components of cyberspace (e. g. a virtual mug) and cyberbody (e.g. virtual hand) as well as the computer-generated objects and real body parts as has been investigated in the previous studies \cite{ambrosecchia2015spatial,costantini2010does,tucker1998relations,tucker2004action}. 

In both Experiment 1 and Experiment 2, we used objects that would potentiate reach-and-grasp action. Even though reaching and grasping seem like complementary actions that would occur in sequence, they are known to be processed at separate parts of the monkey dorsal stream, reaching in the medial intraparietal area \cite{cisek2001embodiment,kalaska1996parietal,snyder2000intention} and grasping in the anterior intraparietal area \cite{jeannerod1995grasping,rizzolatti1998organization}, the human equivalents of which have also been assessed using functional magnetic resonance imaging \cite{grefkes2005functional}. In his affordance competition account \cite{cisek2007cortical}, Cisek  argues that potential actions compete alongside the dorsal stream, and only those that survive this competition are transmitted into the frontal areas such as the dorsal premotor cortex, the activity of which reflects the final action decision \cite{kalaska1995deciding}. In this context, the restricted avatar hand potentiating greater affordance effect than the able avatar hand in our study may be explained by a lack of competing action in the case of restricted hand, as it does not allow any grasping movement but only reaching. This is consistent with recent neurophysiological evidence that when monkeys are provided with two potential reaching targets, action execution is relatively delayed, implying that multiple reach options are all represented at the beginning, and then gradually eliminated until a final decision is made \cite{cisek2005neural}.

Whereas we confirmed the affordance effect during the action planning, as indexed by the lift-off time, the handle orientation effect was still significant but interestingly reversed during the action execution as indicated by the movement time measure. The reverse compatibility effect, which is rare but still existing in the literature \cite{bub2012dynamics,kostov2015reversing,vainio2008relations,yu2014limits}, was explained by a horizontal object positioning, where object centralization could either be made according to the mass or the length of object, the latter of which would yield an uneven distribution of pixels favoring the body \cite{kostov2015reversing}. This would render the handle insignificant, or the body of object containing more task-relevant information than the handle, in either case of which the endogenous attention would shift towards the body and thus, the reaction times would be faster when the body – and not the handle - of a target object is on the same side as the responding hand. In a task, where participants were asked to indicate whether the presented objects are man-made or natural images, Yu et al. have also found a reverse compatibility effect, where the reaction times were faster when the response button was situated on the opposite side as the depicted object's handle \cite{yu2014limits}. Replicating Tucker and Ellis' original stimulus-response compatibility effect in a paradigm which rather required an explicit instruction of imagining picking up each object, Yu et al. concluded that it is not merely observing an object, but rather engaging in an action-relevant motor task which facilitates action for compatible responses.  Similarly, in an effort to explain the negative affordance effect, Vainio drew attention to the presentation time \cite{vainio2011negative}, revealing reversed stimulus-response compatibility during brief presentations of target objects (30 or 70 ms), which turned into a positive compatibility effect when the object was displayed for 370 ms. The authors argued that it takes a build-up time to gradually activate associated motor representations. None of these accounts, however, are adequate in explaining the reversed compatibility effect in our paradigm, as we found both positive and negative stimulus-response compatibility effects using the same stimuli and procedure in the same setup. The discrepancy in our study could rather be originating from two response measures having been collected at different phases of action. Whereas participants started the action faster in the compatible conditions than in the incompatible ones, they rather moved more rapidly in the incompatible conditions than in the compatible ones. This is in agreement with Cisek's argument that action planning is never complete \cite{cisek2007cortical}. Cisek argues that reaching a certain information threshold is enough to start an action. However, ``even in the cases of highly practised behaviours'' (p. 1586), an action begins without a complete trajectory. That is, there is an ongoing cycle of planning and observation during the action execution, with continuous feedback loops that allow fine motor adjustments while interacting with objects. Therefore, actions are better thought of as processes than discrete contained events. Within this theoretical framework, in our experiment, no fine motor adjustments are needed in the conditions where the handle orientation does not correspond to the target location. If the handle is on the same side as the reaching trajectory, however, action planning may still continue to adjust the hand movement to the handle position during the action execution, even though the initial plan may not be to grasp it. This may explain why we found an approximately 14 milliseconds of slow-down in the movement time for the compatible compared to the incompatible conditions.

We rely on the sensory feedback to perceive our bodies and the environment around 
us, which makes it possible for us to control novel bodies in VR \cite{lanier2006homuncular}. The results of Experiment 2 supported the idea of flexibility in body perception. In order to adapt participants to their novel cyber body parts, we introduced a 5-minutes adaptation procedure, where participants were asked to interact with virtual objects in various ways, including pushing, pulling and lifting actions. Our preliminary data demonstrated that this adaptation phase was crucial in obtaining the relevant stimulus-response compatibility effects. It is known that it is possible to embody a virtual hand via synchronous visuo-tactile feedback \cite{aldhous2017digital}. Here, we demonstrated that synchronous visual feedback, even in the absence of haptic input, is enough to create such embodiment effects. 

The purpose of this study was to evaluate the perception of affordances in VR and the effect of avatars on affordance perception. Results indicated that the affordance effect can be created and measured in VR while the participant is interacting with virtual objects. In other words, virtual objects in VR trigger specific actions like real-world objects do. However, it is not only dependent on the properties of the objects. People perceive potential actions in relation to their perceived abilities. We have shown that the representation of the virtual body affects the way people interact with virtual objects. Understanding the effects of embodiment on interactivity in virtual environments is highly important for commercial VR applications to provide better user experience and also for research in computer science, psychology, neuroscience, and cognitive science to better understand human perception through action.  In the final section, we will provide implications for interaction design in VR. 

\section{Implications for interaction design in VR}
Designing interactions in VR is a challenging task. First of all, users need to know what is interact-able and how to interact with it. Just as visual representations of virtual objects drive their use, the visual representation of the body determines the available actions with which objects can be interacted. Thus, changing virtual hands and their capabilities can make it easier for the user to understand potential actions in virtual environments. Altering the body form can help lead users to understand what is possible in a VR setting and thereby improve user experience.

The idea of having shape-shifting hands provides potentially infinite affordances for virtual objects in VR. Through the alteration of avatar abilities, far objects could now become reachable, heavy objects liftable, small objects stretchable or big objects scalable. Enhancing the ways of interaction by giving users magical abilities provides creative freedom for developers, designers, and users. 

Having multiple action possibilities, however, can also distract users, make the application hard to learn and use, and make the decision making harder (as shown by the results of Experiment 2). Therefore, virtual interactions can sometimes be enhanced by restricting potential actions to direct users to perform a task in a certain way. Restricting action possibilities may also provide precise results with imprecise actions. Unlike real object interactions, a user can interact with virtual objects by making actions in the vicinity of the object that is expected to be interacted on. For example, in VR, the detection of contact between any part of the user's hand with a virtual object can be enough to grasp the object. In real life, on the other hand, grasping an object requires a complex sequence of actions and proper positioning of the fingers. This difference may provide advantages in some tasks for VR over the real-world tasks. 

VR provides an environment for the curious to explore the limits of body perception and its effects on interactivity in Human-Computer Interaction. Virtual Reality represents a new frontier for extending representations of the human form, creates new areas of application, and open or even design new perceptions of reality. This will, in turn, create new horizons of research in perception, blurring the lines between the real and the virtual.     

%

%

\begin{acknowledgements}
We provide our thanks to Erhan Oztop, Albert Ali Salah and Esra Mungan
for reading and providing comments to an earlier version of this
manuscript.\newline
Also, we are grateful for the support of Teleporter Realities Inc. which provided the devices used in this study.

\end{acknowledgements}

%
%

\bibliographystyle{spmpsci}      
\bibliography{references}

\newpage

\section{Appendix A}
\centering
\textbf{Appendix A \newline Body Ownership Questionnaire \newline for Able Hands}
\newline

1. I felt as if the virtual hands that I saw during the experiment were part of my body. 
\newline

Strongly Disagree	0    1    2    3    4    5    6    7    8    9   10	Strongly Agree 
\newline				
\newline
\newline

2. I felt as if the virtual hands were moving independently of my movements. 
\newline

Strongly Disagree	0    1    2    3    4    5    6    7    8    9   10	Strongly Agree 
\newline
\newline
\newline

3. I felt as if my body was immersed in the virtual environment.
\newline

Strongly Disagree	0    1    2    3    4    5    6    7    8    9   10	Strongly Agree 
\newline
\newline
\newline

4. I felt as if the virtual hands were someone else’s hands. 					
\newline

Strongly Disagree	0    1    2    3    4    5    6    7    8    9   10	Strongly Agree 
\newline
\newline
\newline

5. I felt as if I was the one who is controlling the virtual hands. 
\newline
Strongly Disagree	0    1    2    3    4    5    6    7    8    9   10	Strongly Agree 
\newline
\newline
\newline

6. During the experiment, I felt as if was watching the scene from a third-person perspective.
\newline

Strongly Disagree	0    1    2    3    4    5    6    7    8    9   10	Strongly Agree

\newpage

\section{Appendix B}
\centering
\textbf{Appendix B \newline Body Ownership Questionnaire \newline for Restricted Hands}
\newline

1. I felt as if the virtual capsule hands that I saw during the experiment were part of my body.
\newline

Strongly Disagree	0    1    2    3    4    5    6    7    8    9   10	Strongly Agree 
\newline				
\newline
\newline

2. I felt as if the virtual capsule hands were moving independently of my movements. 
\newline

Strongly Disagree	0    1    2    3    4    5    6    7    8    9   10	Strongly Agree 
\newline
\newline
\newline

3. I felt as if my body was immersed in the virtual environment. 
\newline

Strongly Disagree	0    1    2    3    4    5    6    7    8    9   10	Strongly Agree 
\newline
\newline
\newline

4. I felt as if the virtual capsule hands were someone else’s hands. 
\newline

Strongly Disagree	0    1    2    3    4    5    6    7    8    9   10	Strongly Agree 
\newline
\newline
\newline

5. I felt as if I was the one who is controlling the virtual capsule hands.
\newline
Strongly Disagree	0    1    2    3    4    5    6    7    8    9   10	Strongly Agree 
\newline
\newline
\newline

6. During the experiment, I felt as if I was watching the scene from a third-person perspective.
\newline

Strongly Disagree	0    1    2    3    4    5    6    7    8    9   10	Strongly Agree

\end{document}